# Magnetism and Transport in YbMn$_2$Sb$_2$

R. Nirmala[1], A. V. Morozkin[2], K. G. Suresh[3], H.-D. Kim[4], J.-Y. Kim[4],
B.-G. Park[5], S.-J. Oh[6] and S. K. Malik[1*],

[1]*Tata Institute of Fundamental Research, Mumbai 400 005, India*
[2]*Department of Chemistry, Moscow Lomonosov State University, Moscow, GSP - 3, 119899, Russia*
[3]*Department of Physics, Indian Institute of Technology, Mumbai 400 076, India*
[4]*Pohang Accelerator Laboratory, Pohang University of Science and Technology, Pohang 790-784, Korea*
[5]*Department of Physics, University of Science and Technology, Pohang 790-784, Korea*
[6]*School of Physics and Center for Strongly Correlated Materials Research, Seoul National University, Seoul 151-742, Korea*

## Abstract

A new ternary intermetallic compound, namely, YbMn$_2$Sb$_2$, has been synthesized and its magnetic and electrical transport properties have been studied in the temperature range of 2 – 300 K. This compound crystallizes in a trigonal, La$_2$O$_2$S type structure (space group P3bm1, No. 164) and is found to be ferromagnetically ordered at room temperature. The magnetism is attributed to the ordering of Mn-sublattice. M$_5$ x-ray absorption spectrum of YbMn$_2$Sb$_2$ obtained at room temperature suggests that the valency of Yb in this compound is close to 2+. Electrical resistivity of this compound is metal-like and a positive magnetoresistance of 13 % is observed at 5 K in an applied field of 9T.

* Corresponding author
E-mail: skm@tifr.res.in

## Introduction

Rare earth intermetallic compounds of the type $RT_2X_2$ (R = Rare earth, T = Transition metal, X = Si, Ge) with Mn at the transition metal site, exhibit exotic magnetic properties [1] because of the competition and interplay of R-R, Mn-Mn and R-Mn exchange interactions. Recently, magnetic properties of $YbMn_2X_2$ (X = Si, Ge) compounds have attracted some attention. The compound $YbMn_2Si_2$ exhibits three magnetic transitions: a collinear antiferromagnetic ordering of Mn moments at 526 K and an antiferromagnetic rearrangement of Mn-sublattice below 30 K followed by the magnetic ordering of $Yb^{3+}$ moments at 1.5 K [2]. The compound $YbMn_2Ge_2$ shows planar antiferromagnetism below 510 K where Mn moments lie in the c-plane with interlayer and intralayer antiferromagnetic coupling. The coupling scheme in this compound changes to non-collinear-type below 185 K, whereas no Yb sublattice magnetic ordering is observed down to 4.2 K [3]. The role of Yb in the magnetic properties of $RMn_2X_2$ compounds is fascinating because recent x-ray photoelectron spectroscopy (XPS) measurements have revealed that Yb is in 3+ state in $YbMn_2Si_2$ but in 2+ state in $YbMn_2Ge_2$ [4]. We have synthesized a new compound, namely, $YbMn_2Sb_2$, and studied its magnetic and electrical transport properties. Replacement of Si/Ge with bigger Sb is likely to lead to a change in Mn-Mn interatomic distances which may result in a novel magnetism in this compound.

## Experimental details

Polycrystalline sample of $YbMn_2Sb_2$ was prepared by arc melting of constituent elements (Yb - 99.9 % pure, Mn – 99.99 % pure, Sb - 99.999 % pure) in argon atmosphere. The sample was turned over and remelted several times. X-ray diffraction data were obtained at room temperature (DRON-3.0 diffractometer, Cu Kα radiation, 2θ = 20° – 70°). Magnetization measurements were carried out in the temperature range of 1.8 K – 330 K using a SQUID magnetometer (MPMS XL, Quantum Design), in applied fields up to 7 T. Magnetization vs. field isotherms at few selected temperatures were obtained in fields up to 12 T in a vibrating sample magnetometer (Oxford Instruments). Electrical resistivity and magnetoresistance measurements were performed by employing linear four-probe technique (PPMS, Quantum Design) in the temperature range of 2 K – 350 K, in applied fields up to 9 T. Differential Scanning Calorimetry (DSC) experiment was performed in the temperature range of 300 K – 600 K (DSC 7, Perkin-Elmer). X-ray absorption spectrum (XAS) of $YbMn_2Sb_2$ compound was collected by a total electron yield method at the 2A1 beamline of Pohang Accelerator Laboratory.

**Results and Discussion**

Rietveld analysis of the room temperature X-ray diffraction data of $YbMn_2Sb_2$ showed that this is a single-phase compound, crystallizing in a trigonal, $La_2O_2S$-type structure (space group P3bm1, No. 164). This is in contrast to the $YbMn_2X_2$ (X = Si, Ge) compounds which form in tetragonal, $ThCr_2Si_2$-type structure (space group I4/mmm, No. 139). The lattice parameters and atomic position parameters of $YbMn_2Sb_2$ are given in Table 1. It may be noted from the table that all the atoms occupy distinct lattice sites.

Magnetization vs. field (M-H) isotherm of $YbMn_2Sb_2$ obtained at 300 K indicates that the compound is ferromagnetically ordered at room temperature (Fig. 1). Since high temperature magnetization measurements could not be performed in our laboratory, differential scanning calorimetry analysis was carried out in the temperature range 300 K – 600 K to determine the Curie temperature. The DSC data reveal a sharp peak at 338 K

(Fig. 2) suggesting the ferromagnetic ordering of the Mn moments at this temperature. This is consistent with the fact that the shortest Mn-Mn distance in $YbMn_2Sb_2$ (3.304 Å) is larger than the critical distance of 2.85 Å required for ferromagnetic ordering of Mn spins to occur.

Magnetization, as a function of temperature, obtained in an applied field of 0.5 T shows a broad hump at 136 K (Fig. 3). This hump temperature gets progressively lowered to 99 K on application of magnetic field of 4 T and to 78 K, in 7 T, respectively (Fig. 3). Similar behaviour has been observed earlier in $YbMn_2Ge_2$ compound where it has been attributed to the reorientation of Mn moments from an initial ferromagnetic to a low temperature canted antiferromagnetic structure [5]. This is also corroborated by M-H isotherms on $YbMn_2Ge_2$ which show a metamagnetic spin-flop transition. It is possible that a similar situation is occurring in the presently studied $YbMn_2Sb_2$ compound. However, the M-H isotherms in this compound obtained at various temperatures indicate a smooth, initial rise in magnetization followed by a linear increase in high fields up to 12 T, with no signs of a metamagnetic transition. Therefore, the hump in magnetization in $YbMn_2Sb_2$ may not correspond to a spin-flop transition. The other possible origin of this hump may be from the mixed valent behaviour of Yb ions. It is known that the magnetic susceptibility of mixed valent Yb compounds shows a broad hump at some temperature [6]. In order to determine the valence state of Yb, we have measured an x-ray absorption spectrum (XAS) of $YbMn_2Sb_2$ compound by a total electron yield method at the 2A1 beamline of Pohang Accelerator Laboratory. Absence of a distinct, peak-like structure around the $M_5$ x-ray edge (figure not shown) indicates a possible 2+ valency and hence a nonmagnetic ground state for Yb. This observation is consistent with low temperature heat capacity measurements (performed down to 1.8 K) on YbMn$_2$Sb$_2$ [7] which neither yield a substantially enhanced value of the electronic specific heat co-efficient, γ, characteristic of mixed-valent systems nor magnetic ordering of the Yb moments. Thus the origin of the hump in the magnetization of YbMn$_2$Sb$_2$ remains unclear and neutron diffraction studies at various temperatures are desired to understand the nature of Mn magnetic ordering in this compound.

Electrical resistivity, ρ(T), of YbMn$_2$Sb$_2$, in the temperature range of 2 K – 350 K, is metallic and is of the order of a few μΩcm (Fig. 4). It varies linearly with temperature as the temperature is lowered and shows a tendency towards saturation, to a constant value, $\rho_0$, at very low temperatures. The magnetic anomalies do not show up in the electrical resistivity data. Magnetoresistance, [MR = ($R_H$-$R_0$) / $R_0$] is about 13 % at 5 K in an applied field of 9 T (inset in Fig. 4) and MR has a positive sign. Increase of MR with magnetic field is quadratic in weak fields and becomes linear at high fields, as expected, for a normal metal.

**Conclusion**

In summary, a new polycrystalline YbMn$_2$Sb$_2$ compound, of trigonal, La$_2$O$_2$S-type structure has been synthesized. The compound orders ferromagnetically at 338 K due to the ordering of Mn moments. The Yb ion appears to be in a near 2+ state and hence non-magnetic in this compound. Humps in the magnetization are observed around 138 K whose origin is not clear as yet.

**Acknowledgements**

The authors thank A.V. Gurjar for performing DSC analysis and S. A. Watpade and D. Buddhikot for their technical help.

**References**

1. T. Shigeoka, N. Iwata, H. Fujii, T. Okamoto, J. Magn. Magn. Mater. 54-57 (1986) 1343.

2. M. Hofmann, S. J. Campbell, A. V. J. Edge and A. J. Studer, J. Phys. Condens. Matter **13** (2001) 9773.

3. T. Fujiwara, H. Fujii, K. Koyama and M. Motokawa, Physica B **312-313** (2002) 864.

4. A. Szytula, A. Jezierski, B. Penc, M. Hofmann and S. J. Campbell, J. Alloys Comp. **366** (2004) 313.

5. I. Nowik, Y. Levi, I. Felner and E. R. Bauminger, J. Magn. Magn. Mater. **147** (1995) 373.

6. See for instance, D. T. Adroja, S. K. Malik, B. D. Padalia, S. N. Bhatia, R. Walia and R. Vijayaraghavan, Phys. Rev. B 42 (1990) 2700.

7. R. Nirmala, A. V. Morozkin and S. K. Malik (to be published).

**Table Captions**

Table 1. Atomic position parameters (x, y, z) of the $La_2O_2S$-type, $YbMn_2Sb_2$ compound [Lattice parameters a = 4.5235 Å and c = 7.4410 Å ; space group P3bm1, No. 164].

**Figure Captions**

Fig. 1. Magnetization vs. Field (M-H) isotherms of $YbMn_2Sb_2$ at 300 K, 100 K and 5 K.

Fig. 2. Differential Scanning Calorimetry data of $YbMn_2Sb_2$ in the temperature range of 300 K to 600 K.

Fig. 3. Magnetization (M) vs. Temperature (T) of $YbMn_2Sb_2$ in various applied fields.

Fig. 4. Electrical resistivity ($\rho$) vs. Temperature (T) of $YbMn_2Sb_2$ in zero field. [Inset: Magnetoresistance (MR%) vs. Field (H) of $YbMn_2Sb_2$ at 5 K]

Table 1. Atomic position parameters (x, y, z) of the trigonal, La$_2$O$_2$S-type, YbMn$_2$Sb$_2$ compound [Lattice parameters a = 4.5235 Å and c = 7.4410 Å ; space group P3bm1, No. 164]

| Atom | x | y | z | Occupancy |
|---|---|---|---|---|
| Yb | 0 | 0 | 0 | 1 |
| Mn | 1/3 | 2/3 | 0.636 | 2 |
| Sb | 1/3 | 2/3 | 0.245 | 2 |

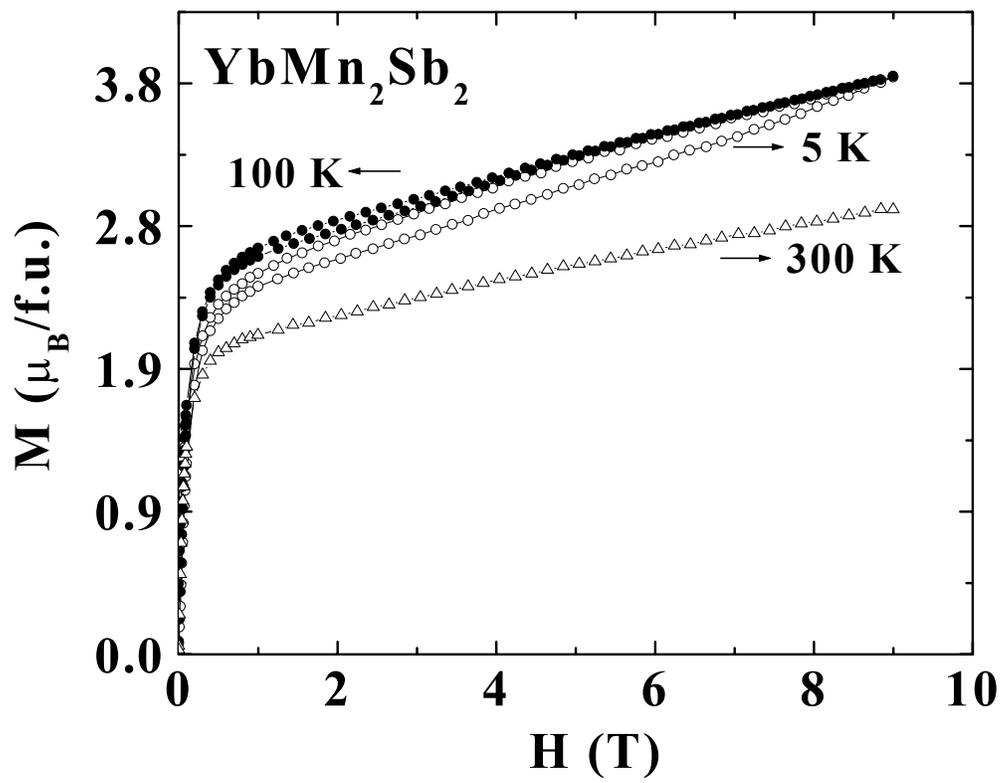

Fig. 1

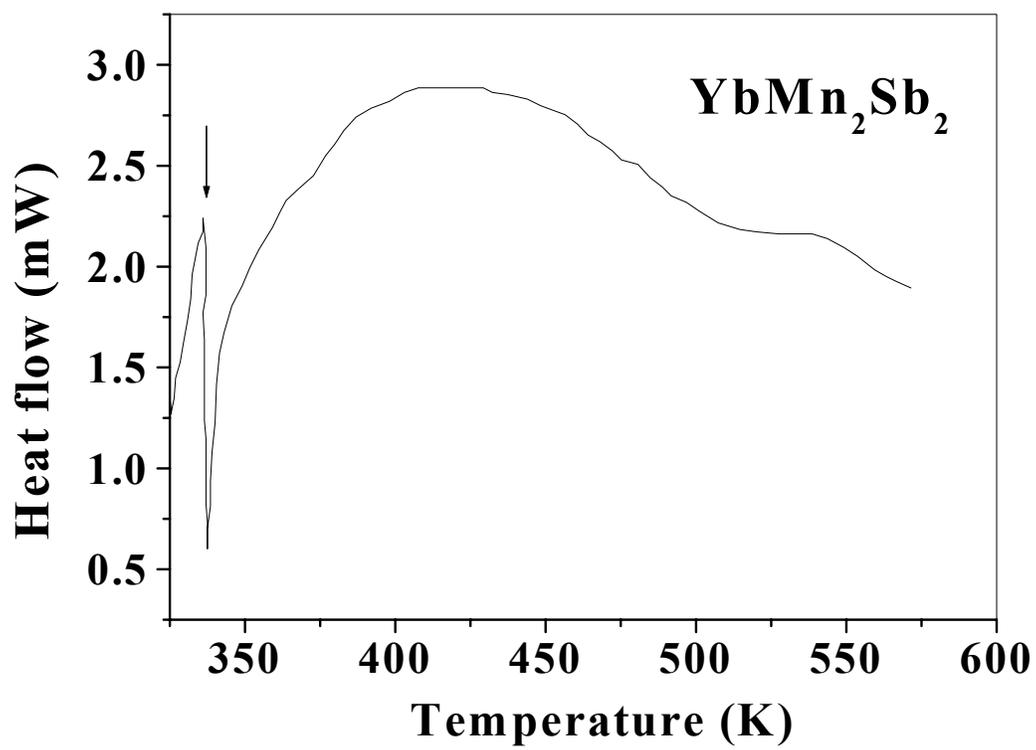

Fig. 2

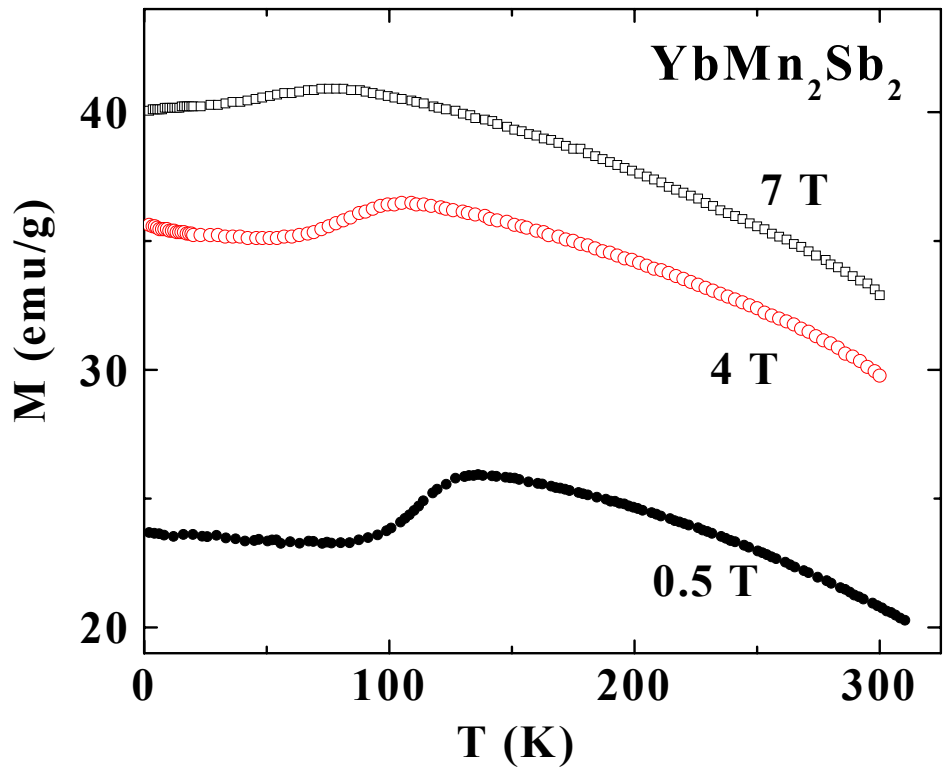

Fig. 3

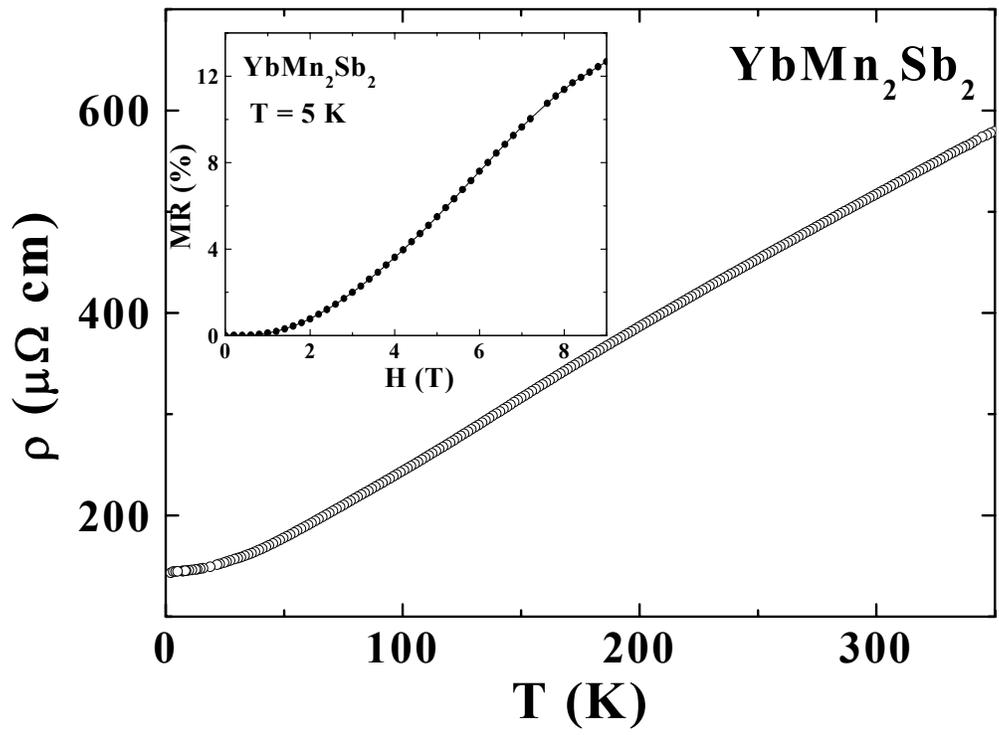

Fig. 4